\documentclass[pre,amsfonts,showpacs,preprint]{revtex4}

\usepackage{calc}
\usepackage{enumerate}
\usepackage{mathtools}
\usepackage{amsmath}
\usepackage{amsthm}
\usepackage{amssymb}
\usepackage{amsbsy}
\usepackage[T1]{fontenc}
\usepackage{textcomp}
\usepackage{ae,aecompl}
\usepackage{txfonts}
\usepackage[colorlinks=true,backref=true,pagebackref=true]{hyperref}
\usepackage{bm}

\newtheorem{thm}{Theorem}
\newtheorem{lem}{Lemma}

\DeclareMathOperator{\diag}{diag}
\DeclareMathOperator{\uni}{uni}
\DeclareMathOperator{\eig}{eig}
\DeclareMathOperator{\vectorize}{vec}
\DeclareMathOperator{\Sym}{{\mathbb{S}}}
\DeclareMathOperator{\SO}{{\it SO}}

\newcommand{\suchthat}{\ensuremath{|\,}}
\newcommand{\rand}[1]{\ensuremath{{\mathcal{#1}}}}

\newcommand{\Real}[1][]{\ensuremath{\mathbb{R}^{#1}}} 

\newtagform{parenthesis}{(}{)}
\usetagform{parenthesis}

\newcommand{\reflem}[2][l]{{#1}emma \ref{#2}}
\newcommand{\refthm}[2][t]{{#1}heorem \ref{#2}}

\newcommand{\function}[5]{\ensuremath{\begin{array}{rrl}
      {#1}:{#2}\negthickspace&\rightarrow&\negthickspace{#3}\\
           {#4}\negthickspace&\mapsto    &\negthickspace{#5}
    \end{array}}}

\begin{document}
\thispagestyle{empty}

\preprint
\date{\today}

\title{Probability Distribution of Curvatures of Isosurfaces in Gaussian Random Fields}

\author{Paulo R.\ S.\ Mendon\c{c}a}
\email{mendonca@crd.ge.com}
\author{Rahul Bhotika}
\email{bhotika@research.ge.com}
\author{James V. Miller}
\email{millerjv@crd.ge.com}
\affiliation{GE Global Research Center,
  Niskayuna, NY 12065, USA}

\begin{abstract}
  An expression for the joint probability distribution of
  the principal curvatures at an arbitrary point in the
  ensemble of isosurfaces defined on isotropic Gaussian
  random fields on $\Real[n]$ is derived. The result is
  obtained by deriving symmetry properties of the ensemble
  of second derivative matrices of isotropic Gaussian random
  fields akin to those of the Gaussian orthogonal ensemble.
\end{abstract}

\pacs{02.50.Ey,02.50.Fz}
\maketitle
\thispagestyle{empty}

\section{Introduction}
We closely follow the notation in \cite{Magnus95} and
\cite{Abrahamsen1997a}.  Let $\mathbf T$ be a tensor with rank
$a$ and dimensions $(d_{1},\ldots,d_{a})$. The bijective
linear map $\vectorize$ associates ${\mathbf T}$ to the vector
$\vectorize{\mathbf T}$ in $\Real[N]$, $N=\prod_{i=1}^{a}d_{i}$,
with entry $(\vectorize{\mathbf T})_{k}$,
$k\in\{1,\ldots,\prod_{i=1}^{a}d_{i}\}$, given by
$(\vectorize{\mathbf T})_{k} = {\mathbf T}_{i_{1},\ldots,i_{a}}$
with $i_{j}\in\{1,\ldots,d_{j}\}$ uniquely defined by
$k=1+\sum_{j=1}^{a}(i_{j}-1)\prod_{k=1}^{j-1}d_{k}$. Let
$d_{i}=n$, $i=1,\ldots,a$.  The linear operator $\diag$ maps
$\mathbf T$ to the vector $\diag{\mathbf T}$ in $\Real[n]$ with
entry $(\diag{\mathbf T})_{k}$, $k\in\{1,\ldots,n\}$, given by
${\mathbf T}_{k,k,\ldots,k}$. If restricted to the set $D$ of
diagonal matrices, $D\ni{\mathbf D}\rightarrow\diag{\mathbf D}$ is
bijective, and therefore the inverse mapping $\diag^{-1}$ is
well-defined. Let $a=2$ and $d_{1}=d_{2}=n$.  The linear
operator $\uni$ maps ${\mathbf T}$ to the vector $\uni{\mathbf
  T}\in\Real[n(n+1)/2]$ with entry $(\vectorize{\mathbf
  T})_{k}$, $k\in\{1,\ldots,n(n+1)/2\}$, given by
$(\vectorize{\mathbf T})_{k} = {\mathbf T}_{i,j}$ uniquely defined
by $k = (j-1)n - j(j-1)/2 + i$. This operator maps a matrix
$\mathbf T$ to a vector containing the entries of $\mathbf T$ read
along its columns, but ignoring the elements above the main
diagonal.

Following the notation in \cite{Magnus95}, the $(n,n)$
identity matrix is denoted by $\mathbf{I}_{n}$, ${\mathbf
  C}_{n}$ denotes the $(n^2,n^2)$ commutation matrix, i.e.,
the unique $(n^2,n^2)$ matrix such that $\vectorize{{\mathbf
    T}^{\rm T}}={\mathbf C}_{n}\vectorize{\mathbf T}$ for all
$(n,n)$ matrices $\mathbf T$. The matrix ${\mathbf D}_{n}$ denotes
the $(n^2,n(n+1)/2)$ duplication matrix, i.e., the unique
$(n^2,n(n+1)/2)$ matrix such that $\vectorize{\mathbf T}={\mathbf
  D}_{n}\uni{\mathbf T}$ for all $(n,n)$ symmetric matrices $\mathbf
T$. From this definition, we have $\uni{\mathbf T}={\mathbf
  D}_{n}^{{+}}\vectorize{\mathbf T}$, where ${\mathbf A}^{+}$
indicates the Moore-Penrose inverse of the matrix ${\mathbf A}$.
The duplication matrix and the commutation matrix are
related through the identity ${\mathbf D}_{n}{\mathbf D}^{{+}}_{n} =
(1/2)({\mathbf I}_{n^2} + {\mathbf C}_{n})$.  Finally, ${\mathbf
  1}_{n,m}$ denotes an $(n,m)$ matrix of ones.

A \emph{random scalar field} in the set $X$ is a stochastic
process, i.e., an indexed set $\rand{F}_{X}=\{\rand{F}_{\mathbf
  x}, {\mathbf x}\in X\}$ of random variables $\rand{F}_{\mathbf x}$
defined over the same probability space
$(\Omega,\sigma_\Omega,P)$.  A \emph{random tensor fields}
is defined analogously, with $\rand{F}_{\mathbf x}$ a
vector-valued function.

Let $\rand{F}_{X}^{1}$ and $\rand{F}_{X}^2$ be two random
scalar fields as above, and assume that $\rand{F}_{\mathbf
  x}^{i}$ is zero mean, which, for the purposes of this
work, implies no loss of generality. If the expectation
$E\{\rand{F}_{{\mathbf x}}^{1}\rand{F}_{{\mathbf y}}^{2}\}$ taken
over the joint distribution of $\rand{F}_{X}^{1}$ and
$\rand{F}_{X}^{2}$ is defined for all ${\mathbf x}$ and ${\mathbf
  y}$ in $X$, the function
$R_{\rand{F}^{1},\rand{F}^{2}}({\mathbf x},{\mathbf y}) =
E\{\rand{F}_{{\mathbf x}}^{1}\rand{F}_{{\mathbf y}}^{2}\}$ defines
the \emph{cross-covariance function} of the random fields.
If $\rand{F}_{X}^1 = \rand{F}_{X}^2 = \rand{F}_{X}$, the
notation is simplified to $R_{\rand{F}} \triangleq
R_{\rand{F},\rand{F}}$, and the function $R_{\rand{F}}$ is
referred to as the \emph{autocovariance function} of
$\rand{F}_{X}$.  The \emph{conditional autocovariance
  function} of $\rand{F}_{X}^1$ given $\rand{F}_{X}^2$,
$R_{\rand{F}^{1}\suchthat \rand{F}^{2}}$, is defined by
taking the expectation of $\rand{F}_{\mathbf x}^1$ over the
conditional distribution of $\rand{F}_{\mathbf x}^1$ and
$\rand{F}_{\mathbf y}^1$ given $\rand{F}_{X}^2$. In the case of
a random tensor field the definitions are analogous, with
the product $\rand{F}_{\mathbf x}^{1}\rand{F}_{\mathbf y}^{2}$
replaced by a tensor product $\vectorize\rand{F}_{\mathbf
  x}^{1}\otimes (\vectorize\rand{F}_{\mathbf y}^{2})^{\rm T}$
and the expectation taken over each entry of the tensor.
Henceforth, the term ``random field'' will be used in
reference to both scalar and tensorial random fields.

Let $X$ be a vector space. Zero-mean random fields
$\rand{F}_{X}^{1}$ and $\rand{F}_{X}^{2}$ on $X$ are
\emph{wide-sense stationary} if their cross-covariance
function $R_{\rand{F}_{X}^{1},\rand{F}_{X}^{2}}({\mathbf x,y})$
satisfies $R_{\rand{F}_{X}^{1},\rand{F}_{X}^{2}}({\mathbf x,y})
= R_{\rand{F}_{X}}({\mathbf s})$ with ${\mathbf s} = {\mathbf x-y}$ for
all ${\mathbf x}$ and ${\mathbf y}$ in $X$. Henceforth we will
assume that $X=\Real[n]$, and therefore the subscript $X$ in
$\rand{F}_{X}$ can then be safely omitted by defining
$\rand{F}\triangleq\rand{F}_{\Real[n]}$.  A stationary
random field on $\Real[n]$ is \emph{isotropic} if its
autocovariance function $R_{\rand{F}}({\mathbf s})$ satisfies
$R_{\rand{F}}({\boldsymbol{s}}) = \sigma^2\rho(\lVert{\mathbf
  s}\rVert)$, where $\rho$ is a correlation function
\cite{Abrahamsen1997a} and $\lVert\cdot\rVert$ is the
standard Euclidean norm in $\Real[n]$.

\section{Derivatives of Isotropic Random
  Fields\label{sec:derivativesOfRandomFields}}
Great simplification is achieved in the derivations that
follow if $\rho(\lVert{\mathbf s}\rVert)$ can be rewritten as
$\rho(\lVert{\mathbf s}\rVert) = r(\lVert{\mathbf
  s}\rVert^2)/\sigma^2 \Leftrightarrow \rho(\sqrt{\lVert{\mathbf
    s}\rVert}) = r(\lVert{\mathbf s}\rVert)/\sigma^2$. For
$\lVert{\mathbf s}\rVert > 0$ the smoothness of $r(\lVert{\mathbf
  s}\rVert)$ is contingent upon that of $\rho(\lVert{\mathbf
  s}\rVert)$. However, for $\lVert{\mathbf s}\rVert=0$ the
non-differentiability of $\sqrt{\lVert{\mathbf s}\rVert}$ could
be a problem.  This is not the case, as shown in appendix.
The symbols $\rho_{0}^{(i)}$
and $r_{0}^{(i)}$ denotes the $i$-th derivative of
$\rho(\lVert{\mathbf s}\rVert)$ and $r(\lVert{\mathbf s}\rVert)$
with respect to ${s}=\lVert{\mathbf s}\rVert$ at ${s}=0$.

Let $f:\Real[n]\rightarrow\Real$ be a scalar function. The
symbol $\frac{\partial^{a+b} f}{\partial{\mathbf x}^{a\rm
    T}\partial{\mathbf x}^b}$ is used to describe the matrix of
dimensions $(n^a,n^b)$ of partial derivatives of $f$, i.e,
\begin{equation}
  \left(\frac{\partial^{a+b} f}{\partial{\mathbf x}^{a\rm T}\partial{\mathbf
        x}^b}\right)_{A,B} \triangleq 
  \frac{\partial^{a+b}f}{\partial x_{a_1}\ldots\partial
    x_{a_a}\partial x_{b_1}\ldots\partial x_{b_b}},\notag
\end{equation}
where $a_{i}\in\{1,2,\ldots,n\}$ and
$b_{j}\in\{1,2,\ldots,n\}$ are uniquely defined by $A =
1+\sum_{i=1}^{n}(a_{i}-1)n^{i-1}$ and $B =
1+\sum_{j=1}^{n}(b_{j}-1)n^{j-1}$.  Second differentiability
of the autocovariance function $R_{\rand{F}}({\mathbf x},{\mathbf
  y})$ of a random field at the pair $({\mathbf x,x})$ implies
\emph{mean square differentiability} of the random field
itself, as demonstrated in theorem 2.4 of
\cite{Abrahamsen1997a}. If a mean-square differentiable
random field is stationary, its derivatives will also be
stationary.  For $\rand{F}^{1}$ and $\rand{F}^{2}$ jointly
stationary with cross-covariance
$R_{\rand{F}^{1},\rand{F}^{2}}({\mathbf s})$, we define
$R_{\rand{F}^{1},\rand{F}^{2}}^{0}\triangleq
R_{\rand{F}^{1},\rand{F}^{2}}({\mathbf 0})$.

The theorem that follows is central to this work:
\begin{thm} \label{thm:covarianceMatrices} Let $\rand{F}$ be
  an isotropic random field on $\Real[n]$ with
  autocovariance function $R_{\rand{F}}({\boldsymbol{s}})=
  \sigma^2\rho(\lVert{\mathbf s}\rVert)$ where
  $\rho({{s}})$ is four-differentiable. 
  Let $\partial \rand{F}$ and $\partial^2\rand{F}$ be the
  tensor fields defined as
  \begin{equation}
    \partial\rand{F}  \triangleq \frac{\partial\rand{F}}{\partial {\mathbf
        x}^{\rm T}}\quad\text{and}\quad
    \partial^2\rand{F} \triangleq
    \frac{\partial^2\rand{F}}{\partial{\mathbf x}^{\rm
        T}\partial{\mathbf x}}.\notag
  \end{equation}
  Then
  \begin{subequations}
    \begin{align}
      R_{\partial\rand{F}}^{0} &
      = -\sigma^{2}{\rho}^{(2)}_{0}\mathbf{I}_{n},\label{eq:covDF}\\
      R_{\partial\rand{F}\!,\partial^2\rand{F}}^{0} & =
      \mathbf{O}_{n,n^2},\label{eq:independence}\\
      R_{\partial^2\rand{F}}^{0} & =
      \sigma^{2}{\rho}^{(4)}_{0}(\mathbf{I}_{n^2}+
      \mathbf{C}_{n}+
      \vectorize\mathbf{I}_{n}\vectorize^{\rm T}\mathbf{I}_{n}).\label{eq:covD2F}
    \end{align}
  \end{subequations}
\end{thm}
\begin{proof}
  Using \reflem{lem:differentiability} in appendix
  \ref{apen:differentiability} we write $\rho(\lVert{\mathbf
    s}\rVert) = r(\lVert{\mathbf s}\rVert^2)/\sigma^2$. From the
  four-differentiability of $r$ and theorem 2.4 in
  \cite{Abrahamsen1997a} we have
  \begin{subequations}
    \begin{align}
      R_{\partial\rand{F}}({\boldsymbol{s}}) & =
      -\frac{\partial^2 r(\lVert{\mathbf s}\rVert^2)}
      {\partial{\boldsymbol{s}}^{\rm T}\partial{\boldsymbol{s}}},\label{eq:R_DF}\\
      R_{\partial\rand{F},\partial^2\rand{F}}({\boldsymbol{s}})
      & = -\frac{\partial^3
        r(\lVert{\mathbf s}\rVert^2)}{\partial{\boldsymbol{s}}^{\rm
          T}\partial{\boldsymbol{s}}^{2}},\label{eq:R_DFD2F}\\
      R_{\partial^2\rand{F}}({\boldsymbol{s}}) & = \frac{\partial^4
        r(\lVert{\mathbf s}\rVert^2)}{\partial{\boldsymbol{s}}^{2\rm
          T}\partial{\boldsymbol{s}}^{2}}.\label{eq:R_D2F}
    \end{align}
  \end{subequations}
  The chain rule can be used to expand 
  \eqref{eq:R_DF}--\eqref{eq:R_D2F}, and
  substituting the identities $\frac{\partial{\mathbf x}}{\partial{\mathbf x}} = \frac{\partial{\mathbf
      x}^{\rm T}}{\partial{\mathbf x}^{\rm T}} = \mathbf{I}_{n}$,
  $\frac{\partial({\mathbf x}^{\rm T}\otimes\mathbf{I}_{n})}{\partial{\mathbf x}^{\rm T}} =
  \mathbf{I}_{n}\otimes\mathbf{I}_{n}$ and
  $\frac{\partial(\mathbf{I}_{n}\otimes{\mathbf x}^{\rm T})}{\partial{\mathbf x}^{\rm T}} =
  \mathbf{C}_{n}$ in the result produces
  \begin{subequations}
    \begin{alignat}{1}
      R_{\partial\rand{F}}({\boldsymbol{s}}) & = 
      {-4}r^{(2)}(\lVert{\mathbf s}\rVert^2){\boldsymbol{s}}\otimes{\boldsymbol{s}}^{\rm T} - 
      2{r}^{(1)}(\lVert{\mathbf s}\rVert^2)\mathbf{I}_{n},\notag\\
      R_{\partial\rand{F},\partial^2\rand{F}}({\boldsymbol{s}}) & = 
      {-4}\big\{2r^{(3)}(\lVert{\mathbf s}\rVert^2){\boldsymbol{s}}\otimes{\boldsymbol{s}}^{\rm
        T}\otimes{\boldsymbol{s}}^{\rm T} + \notag\\
      &\makebox[\widthof{\ensuremath{\displaystyle{} = {}}}]{}
      r^{(2)}(\lVert{\mathbf s}\rVert^2)({\boldsymbol{s}}^{\rm
        T}\otimes\mathbf{I}_{n}+\notag\\
      &\makebox[\widthof{\ensuremath{\displaystyle{} = {}}}]{}
      \mathbf{I}_{n}\otimes{\boldsymbol{s}}^{\rm T} +
      {\boldsymbol{s}}\otimes\vectorize^{\rm
        T}\mathbf{I}_{n})\big\},
      \notag\\
      R_{\partial^2\rand{F}}({\boldsymbol{s}}) & =
      4\big\{8r^{(4)}(\lVert{\mathbf s}\rVert^2){\boldsymbol{s}}\otimes{\boldsymbol{s}}
      \otimes{\boldsymbol{s}}^{\rm
        T}\otimes{\boldsymbol{s}}^{\rm
        T} + \notag\\
      &\makebox[\widthof{\ensuremath{\displaystyle{} = {}}}]{}
      8r^{(3)}(\lVert{\mathbf s}\rVert^2) \big(
      \vectorize\mathbf{I}_{n}\otimes{\boldsymbol{s}}^{\rm T}\otimes{\boldsymbol{s}}^{\rm T} + \notag\\
      &\makebox[\widthof{\ensuremath{\displaystyle{} = {}}}]{}
      2{\boldsymbol{s}}\otimes(\mathbf{I}_{n}\otimes{\boldsymbol{s}}^{\rm
        T} + {\boldsymbol{s}}^{\rm T}\otimes\mathbf{I}_{n}) +
      {\boldsymbol{s}}\otimes{\boldsymbol{s}}\otimes\vectorize\mathbf{I}_{n}^{\rm T}
      \big) + \notag\\
      &\makebox[\widthof{\ensuremath{\displaystyle{} = {}}}]{}
      r^{(2)}(\lVert{\mathbf s}\rVert^2)
      (\mathbf{I}_{n}\otimes\mathbf{I}_{n} + \mathbf{C}_{n} + 
      \vectorize\mathbf{I}_{n}\otimes\vectorize^{\rm
        T}\mathbf{I}_{n})\big\}.\notag
    \end{alignat}
  \end{subequations}
  Making ${\boldsymbol{s}} = {\mathbf 0}$
  completes the proof.
\end{proof}

\section{Curvatures of Gaussian Random Fields}

Let $f({\mathbf x})$ be a second-differentiable scalar function
on $\Real[n]$. For each ${\mathbf x}\in\Real[n]$ for which
$\partial f/\partial{\mathbf x}\neq\mathbf 0$ we define the set
$F_{\mathbf x}$ as $F_{\mathbf x} = \{{\mathbf y}\in\Real[n]\text{ such
  that }f({\mathbf y})=f({\mathbf x})\text{ and }\partial
f/\partial{\mathbf y}\neq\mathbf 0\}$. If ${F}_{\mathbf
  x}\neq\varnothing$, ${F}_{\mathbf x}$ is a
$(n-1)$-hypersurface in $\Real[n]$ \cite{ONeill66}. Let
$\partial f \triangleq \partial f/\partial{\mathbf x}^{\rm T}$
and $\partial^2f \triangleq \partial^2 f/\partial{\mathbf
  x}^{\rm T}\partial{\mathbf x}$.  The \emph{principal
  curvatures} of the hypersurface ${F}_{\mathbf x}$ at $\mathbf x$
are given by the set of eigenvalues $\kappa$ obtained by
solving the eigenproblem
\begin{equation}
-\Bigg(\mathbb{I}_{n} - \frac{\partial f\otimes(\partial f)^{\rm T}}
{\lVert\partial f\rVert^2}\Bigg)\frac{\partial^2f}{\lVert\partial
  f\rVert}\Bigg|_{\mathbf x}{\mathbf
  v}=\kappa{\mathbf v},\label{eq:SpivakEquation} 
\end{equation}
with ${\mathbf v}\in\Real[n]$, $\lVert{\mathbf v}\rVert = 1$ and ${\mathbf v}^{\rm
  T}\partial f = 0$ \cite[pg.\ 138]{Spivak1999III}.
Let $\{{\mathbf n}_{i}, i=1,\ldots,n-1\}$ be an orthornormal basis for the null-space
of $\partial f$, i.e., ${\mathbf n}_{i}^{\rm T}\partial f = 0$ and ${\mathbf n}_{i}^{\rm
  T}{\mathbf n}_{j} = \delta_{ij}$, and let ${\mathbf N}$ be the matrix
${\mathbf N} = [{\mathbf n}_{1}\dots{\mathbf n}_{n-1}].$
The eigenproblem in \eqref{eq:SpivakEquation} can be rewritten as
\begin{equation}
-\frac{{\mathbf N}^{\rm T}(\partial^2f){\mathbf N}}{\lVert\partial
  f\rVert}\Bigg|_{\mathbf x}{\mathbf u} =
\kappa{\mathbf u},\label{eq:spivakEquation}
\end{equation}
with ${\mathbf u}\in\Real[n-1]$, $\lVert{\mathbf u}\rVert = 1$. Equation
\eqref{eq:spivakEquation} is still valid if the function $f$
is the realization $\rand{F}(\omega)$, $\omega\in\Omega$, of
a mean-square second-differentiable scalar random field
$\rand{F}$ on $\Real[n]$. Therefore the random tensor field
$\rand{K}$ of curvatures of isotropic Gaussian random fields
is implicitly defined at $\mathbf x$ such that
$\partial\rand{F}_{\mathbf x}(\omega)\neq\mathbf 0$ by the solutions
of the equation
\begin{equation}
  -\frac{{\rand{N}}^{\rm T}(\partial^2\rand{F}){\mathbf
      \rand{N}}}{\lVert\partial\rand{F}\rVert}{\rand{U}} =
  \rand{K}\rand{U},\label{eq:simplifiedSpivakEquation}
\end{equation}
where $\rand{N}$ is a random tensor field satisfying
$\rand{N}_{\mathbf x}^{\rm T}\partial\rand{F}_{\mathbf x} = 0$, and
$\rand{N}_{\mathbf x}^{\rm T}\rand{N}_{\mathbf x} = {\mathbf I}_{n-1}$,
and $\rand{U}$ is the tensor field such that $\rand{U}_{\mathbf
  x}(\omega)$ are the eigenvectors associated to the
eigenvalues $\rand{K}_{\mathbf x}(\omega)$. 

Henceforth we assume that $\rand{F}$ is an isotropic,
second-differentiable \emph{Gaussian random field}, which is
defined simply as an isotropic random field for which the
joint distribution of any finite set of random variables
$\{\rand{F}_{A}\}$, $A\in\Real[n]$, is Gaussian.  This
assumption implies that the zero-mean random tensors
$\partial\rand{F}$ and $\partial^{2}\rand{F}$ are also
Gaussian, and therefore $\partial\rand{F}_{\mathbf x}$ and
$\partial^{2}\rand{F}_{\mathbf x}$ are fully characterized by
their covariance matrices, given by \eqref{eq:covDF} and
\eqref{eq:covD2F} in \refthm{thm:covarianceMatrices}.
However, because $\partial^{2}\rand{F}$ is symmetric, its
probability density must be handled with care, since
$R_{\partial^{2}\rand{F}}^{0}$ is not invertible.

The following lemma is a trivial corollary of the theorems in
\cite[sec. 7]{Magnus95}.
\begin{lem} \label{lem:matrixIdentity}Let ${\mathbf A}_{n}$ be a $(n,n)$ invertible
  matrix. Then
\begin{equation}
  ({\mathbf A}_{n}\otimes{\mathbf
    A}_{n})R_{\partial^2\rand{F}}^{0}({\mathbf
    A}_{n}^{-1}\otimes{\mathbf A}_{n}^{-1}) = 
  R_{\partial^2\rand{F}}^{0}.
\end{equation}
\end{lem}

Let $\partial^{2}\rand{F}$ be as in 
\refthm{thm:covarianceMatrices}, and let $\rand{R}$ be a
$(n,m)$, $n\geq m$, orthonormal tensor field independent of
$\partial^{2}\rand{F}$, i.e., a random tensor field such
that for all ${\mathbf x}\in\Real[n]$ any realization
$\rand{R}(\omega)$ of $\rand{R}$ satisfies $\rand{R}_{\mathbf
  x}^{\rm T}(\omega)\rand{R}_{\mathbf x}(\omega) = {\mathbf I}_{m}$
and $\rand{R}_{\mathbf x}$ is independent of
$\partial^{2}\rand{F}_{\mathbf x}$.  Define
$\partial^{2}\rand{F}' \triangleq {\rand{R}}^{\rm
  T}\partial^{2}\rand{F}{\rand{R}} = \{ {\rand{R}_{\mathbf
    x}}^{\rm T}\partial^{2}\rand{F}_{\mathbf x}{\rand{R}_{\mathbf
    x}},{\mathbf x}\in\Real[n]\}$.  We prove the following
lemma:
\begin{lem} \label{lem:invariance} 
  $\partial^{2}\rand{F}'$ is a Gaussian random tensor field
  with autocovariance function
  $R_{\partial^{2}\rand{F}'}({{\mathbf s}})$ such that
\begin{align}
  R_{\partial^{2}\rand{F}'}^{0} & =
  \sigma^{2}{\rho}^{(4)}_{0}(\mathbf{I}_{m^2}+\mathbf{C}_{m}+
  \vectorize\mathbf{I}_{m}\vectorize^{\rm T}\mathbf{I}_{m})\label{eq:covarianceN}.
\end{align}
\end{lem}

\begin{proof} 
  Let $P_{\partial^{2}\rand{F}_{\mathbf x}'\suchthat \rand{R}_{\mathbf x}}$ be
  the conditional probability of the random tensor
  $\partial^{2}\rand{F}_{\mathbf x}'$ given $\rand{R}_{\mathbf x}$.
  Since $\partial^{2}\rand{F}_{\mathbf x}$ is independent of
  $\rand{R}_{\mathbf x}$, $\partial^{2}\rand{F}_{\mathbf x}'$ given
  $\rand{R}_{\mathbf x}$ is a linear function of
  $\partial^{2}\rand{F}_{\mathbf x}$, and therefore it is
  zero-mean Gaussian.  Using the identity $\vectorize({\mathbf
    ABC})=({\mathbf C}^{\rm T}\otimes{\mathbf A})\vectorize{\mathbf B}$
  and properties of commutator matrices \cite{Magnus95} we
  can write $\vectorize\partial^{2}\rand{F}_{\mathbf x}' =
  ({\rand{R}_{\mathbf x}}^{\rm T}\otimes{\rand{R}_{\mathbf x}}^{\rm
    T})\vectorize\partial^{2}\rand{F}_{\mathbf x}$, and therefore
\begin{align}
  R_{\partial^{2}\rand{F}_{\mathbf x}'\suchthat {\rand{R}_{\mathbf x}}}^{0} & =
  ({\rand{R}_{\mathbf x}}^{\rm T}\otimes{\rand{R}_{\mathbf x}}^{\rm T})
  R_{\partial^{2}\rand{F}}^{0}({\rand{R}_{\mathbf
      x}}\otimes{\rand{R}_{\mathbf x}})\\
  & =
  \sigma^{2}{\rho}^{(4)}_{0}(\mathbf{I}_{m^2}+
  \mathbf{C}_{m}+\vectorize\mathbf{I}_{m}\vectorize^{\rm
    T}\mathbf{I}_{m}), \label{eq:covarianceNConditional}
\end{align}
using \reflem{lem:matrixIdentity}.  Since
$\partial^{2}\rand{F}_{\mathbf x}'$ given $\rand{R}_{\mathbf x}$ is
zero-mean Gaussian and $R_{\partial^{2}\rand{F}'\suchthat
  {\rand{R}_{\mathbf x}}}^{0}$ does not depend on $\rand{R}_{\mathbf x}$ for
fixed $m$ and $n$, we have $P_{\partial^{2}\rand{F}_{\mathbf
    x}'\suchthat \rand{R}_{\mathbf x}} =
P_{\partial^{2}\rand{F}_{\mathbf x}'}$.  Therefore
$P_{\partial^{2}\rand{F}_{\mathbf x}'}$ is zero-mean Gaussian
with autocovariance function satisfying
\eqref{eq:covarianceN}.
\end{proof}
\reflem[L]{lem:invariance} justifies the notation
$\partial^2\rand{F}^{m}\triangleq\rand{R}^{\rm
  T}\partial^2\rand{F}\rand{R}$, for $\rand{R}$ $(n,m)$.
Let ${\delta^2\rand{F}^{n}}$ be the $(n(n+1)/2,1)$
vector defined as ${\delta^2\rand{F}^{n}} \triangleq
\uni{{\partial^2\rand{F}^{n}}} = {\mathbf
  D}_{n}^{{+}}\vectorize{\partial^2\rand{F}^{n}}$.  Its
covariance matrix $\boldsymbol{\Sigma}_{n}$ is invertible
and given by
\begin{equation}
\boldsymbol{\Sigma}_{n} = {\mathbf
  D}_{n}^{{+}}R_{\partial^2\rand{F}^{n}}^{0}{\mathbf
  D}_{n}^{{+},\rm T}. 
\label{eq:Sigma}
\end{equation}
Therefore the probability
density $p_{\partial^2\rand{F}^{n}_{\mathbf x}}$ of
$\partial^2\rand{F}^{n}_{\mathbf x}$ is standard:
\begin{equation}
  p_{{\delta^2\rand{F}^{n}_{\mathbf x}}}({\mathbf h}) = 
  \frac{1}{\sqrt{\lvert 2\pi{\boldsymbol{\Sigma}_{n}}\rvert}}\exp\bigg(-\frac{{\mathbf h}^{\rm
      T}\boldsymbol{\Sigma}_{n}^{-1}{\mathbf h}}{2}\bigg).
\end{equation}
We now define the random fields
$(\rand{R}^{n},\rand{L}^{n}) \triangleq\{(\rand{R}^{n}_{\mathbf
  x},\rand{L}^{n}_{\mathbf x})\in\SO(n)\times\Real[n]$, ${\mathbf
  x}\in\Real[n]\suchthat$ $\rand{R}_{\mathbf x}^{n,\rm
  T}\rand{R}_{\mathbf x}^{n}={\mathbf I}_{n}$ and $\rand{R}_{\mathbf
  x}^{n,\rm T}\diag^{-1}\rand{L}^{n}_{\mathbf x}\rand{R}_{\mathbf x}^{n} =
\partial^2\rand{F}_{\mathbf x}^{n}\}$,  where $\SO(n)$ is the
special orthogonal group of $(n,n)$ matrices. Let
$\eig^{-1}_{n}$ be the mapping given by
\begin{equation}
  \function{\eig^{-1}_{n}}{\SO(n)\times\Real[n]}{\Sym(n)}
  {({\mathbf R},{\boldsymbol{\Lambda}})}{{\mathbf S}=\eig^{-1}_{n}({\mathbf
      R},{\boldsymbol{\Lambda}})},
\end{equation}
where $\Sym(n)$ is the set of $(n,n)$ symmetric matrices.
This mapping is differentiable and onto, and therefore the
joint probability density $p_{\rand{R}_{\mathbf x},\rand{L}_{\mathbf
    x}}$ of $\rand{R}_{\mathbf x}$ and $\rand{L}_{\mathbf x}$ is
given by
\begin{equation}
  p_{\rand{R}_{\mathbf x}^{n},\rand{L}_{\mathbf x}^{n}}({\mathbf
    R},\boldsymbol{\lambda}) =
  p_{\partial^2\rand{F}^{n}_{\mathbf x}}
  (\uni({{\mathbf R}^{\rm T}\diag^{-1}\boldsymbol{\lambda}{\mathbf R}})) 
  \lvert J({\mathbf R},\boldsymbol{\lambda})\rvert,
  \label{eq:jointRandL}
\end{equation}
$\lvert J({\mathbf R},\boldsymbol{\lambda})\rvert$ is the
absolute value of the Jacobian determinant of
$\eig^{-1}_{n}$.

Theorem 3.3.1 in \cite{MehtaLM2004} provides a
``closed-form'' expression of the probability density of the
eigenvalues of random matrices in the \emph{Gaussian
  orthogonal ensemble} ($\text{GOE}_{n}$). This is the
ensemble of $(n,n)$ real symmetric matrices $\rand{M}$ with
probability density invariant with respect to similarity
transformations $\rand{M}\rightarrow \mathbf R^{\rm T}\rand{M}R$
for any given $(n,n)$ orthonormal $\mathbf R$ \emph{and} such
that the probability distribution of distinct entries are
independent from each other.  Even though each realization
of $\partial^2\rand{F}^{n}$ is real, symmetric, and, by
applying \reflem{lem:invariance}, invariant to the same
similarity transformations, $\partial^2\rand{F}^{n}$ is
different from $\text{GOE}_{n}$, because the distinct
entries of $\partial\rand{F}_{\mathbf x}^{n}$ are not
independent.  However, the assumption of independency used
in \cite{MehtaLM2004} is important only to derive an
expression for the joint probability density of the entries
of random matrices in $\text{GOE}_{n}$, and we already have
that for the matrices in ${\delta^2\rand{F}^{n}}$. Once such
an expression is available the result in \cite{MehtaLM2004}
derives from the observation that, in an expression
analogous to \eqref{eq:jointRandL}, the term $\mathbf R$
appeared only on $\lvert J({\mathbf
  R},\boldsymbol{\lambda})\rvert$, and therefore the
probability density of the eigenvalues of matrices in
$\text{GOE}_{n}$ is obtained through the integration of
$\lvert J({\mathbf R},\boldsymbol{\lambda})\rvert$ over
$\SO(n)$. The next lemma shows that this is also the case
for the probability density $p_{\rand{L}^{n}_{\mathbf x}}$ of
the eigenvalues of $\partial^2\rand{F}^{n}_{\mathbf x}$:
\begin{lem} \label{lem:independence} Let
  $\boldsymbol{\lambda}\in\Real[n]$, ${\mathbf R}\in\SO(n)$, and
  $\boldsymbol{\Sigma}_{n}$ as in \eqref{eq:Sigma}.  Then
  \begin{equation}
    p_{\delta^2\rand{F}^{n}_{\mathbf x}} (\uni({\mathbf R}^{\rm
      T}\diag^{-1}\boldsymbol{\lambda}{\mathbf R})) =
    \frac{1}{\sqrt{\lvert 2\pi{\boldsymbol{\Sigma}_{n}}\rvert}}\exp\bigg(-\frac{{\boldsymbol{\lambda}}^{\rm
        T}\tilde{\boldsymbol{\Sigma}}_{n}^{-1}{\boldsymbol{\lambda}}}{2}\bigg),
  \end{equation}
  where $\tilde{\boldsymbol{\Sigma}}_{n} = {\sigma^{2}{\rho}^{(4)}_{0}}(2\mathbf{I}_{n}+{\mathbf
    1}_{n,1}{\mathbf 1}^{\rm T}_{n,1})$.
\end{lem}
\begin{proof} 
  The following identity can be easily verified:
  \begin{align}
    ({R_{\partial^2\rand{F}^{n}}^{0}})^{{+}} & = 
    \frac{1}{4\sigma^{2}{\rho}^{(4)}_{0}}\bigg(\mathbf{I}_{n^2}
    +\mathbf{C}_{n}-
    \frac{2}{2+n}\vectorize\mathbf{I}_{n}\vectorize^{\rm
      T}\mathbf{I}_{n}\bigg).\label{eq:pinvR}
  \end{align}
  Let $\boldsymbol{\Lambda}=\diag^{-1}\boldsymbol{\lambda}$
  and ${\mathbf u}_{\mathbf R} \triangleq \uni({\mathbf R}^{\rm
    T}\boldsymbol{\Lambda}{\mathbf R})$.  Therefore,
  \begin{align}
    {\mathbf u_R}^{\rm T}\boldsymbol{\Sigma}_{n}^{-1}{\mathbf u_R} &
    = [\uni({\mathbf R}^{\rm T}\boldsymbol{\Lambda}{\mathbf
      R})]^{\rm T}\boldsymbol{\Sigma}_{n}^{-1}[\uni({\mathbf
      R}^{\rm
      T}\boldsymbol{\Lambda}{\mathbf R})]\notag\\
    & = [{\mathbf D}^{{+}}_{n} \vectorize({\mathbf R}^{\rm T}
    \boldsymbol{\Lambda}{\mathbf R}) ]^{\rm
      T}\boldsymbol{\Sigma}_{n}^{-1} [{\mathbf D}^{{+}}_{n}
    \vectorize({\mathbf R}^{\rm T}
    {\boldsymbol{\Lambda}}{\mathbf R})]\notag\\
    & = (\vectorize\boldsymbol{\Lambda})^{\rm T} ({\mathbf
      R}\otimes{\mathbf R}){\mathbf D}_{n}^{{+},\rm T}
    \boldsymbol{\Sigma}_{n}^{-1} {\mathbf D}_{n}^{{+}} ({\mathbf
      R}^{\rm T}\otimes{\mathbf R}^{\rm T})
    \vectorize\boldsymbol{\Lambda},\notag\\
    \shortintertext{and, using $[({\mathbf R}\otimes{\mathbf R}){\mathbf
        D}_{n}]^{{+}} = {\mathbf D}_{n}^{{+}}({\mathbf R}^{\rm
        T}\otimes{\mathbf R}^{\rm T})$,} 
    {\mathbf u_R}^{\rm
      T}\boldsymbol{\Sigma}_{n}^{-1}{\mathbf u_R} & =
    (\vectorize\boldsymbol{\Lambda})^{\rm T} [({\mathbf
      R}\otimes{\mathbf R}){\mathbf D}_{n} \boldsymbol{\Sigma}_{n}
    {\mathbf D}_{n}^{\rm T} ({\mathbf R}^{\rm T}\otimes{\mathbf R}^{\rm
      T})]^{{+}}
    \vectorize\boldsymbol{\Lambda},\notag\\
    \shortintertext{which, using ${\mathbf
        D}_{n}\boldsymbol{\Sigma}_{n}{\mathbf D}_{n}^{\rm T} =
      R_{\partial^2\rand{F}^{n}}^{0}$, yields} {\mathbf
      u_R}^{\rm T}\boldsymbol{\Sigma}_{n}^{-1}{\mathbf u_R} & =
    (\vectorize\boldsymbol{\Lambda})^{\rm T} [({\mathbf
      R}\otimes{\mathbf R}){R_{\partial^2\rand{F}^{n}}^{0}}
    ({\mathbf R}^{\rm T}\otimes{\mathbf R}^{\rm T})]^{{+}}
    \vectorize\boldsymbol{\Lambda}\notag\\
    & = (\vectorize\boldsymbol{\Lambda})^{\rm T}
    ({R_{\partial^2\rand{F}^{n}}^{0}})^{{+}}
    \vectorize\boldsymbol{\Lambda},\notag\\
    & = \boldsymbol{\lambda}^{\rm
      T}\tilde{\boldsymbol{\Sigma}}_{n}^{-1}
    \boldsymbol{\lambda}
    \label{eq:nearlyThere}
  \end{align}
  using \reflem{lem:matrixIdentity} and \eqref{eq:pinvR}.
\end{proof}

The integration of $\lvert J({\mathbf
  R},\boldsymbol{\lambda})\rvert$ over $\SO(n)$, carried out
in \cite{MehtaLM2004}, gives
\begin{equation}
  \int_{\SO(n)}\lvert J({\mathbf
    R},\boldsymbol{\lambda})\rvert\,d{\mathbf R} = 
  \bigg(\frac{\pi^{{(n+1)}/{4}}}{2}\bigg)^{n}
  \frac{\prod_{i=1}^{n-1}\prod_{j=i+1}^{n}\lvert\lambda_{j}-\lambda_{i}\rvert}
  {{\prod_{i=1}^{n}\Gamma(1+{i}/{2})}},
\end{equation}
and it can be shown that the determinant of
$\boldsymbol{\Sigma}_{n}$ is given by
\begin{equation}
\lvert\boldsymbol{\Sigma_{n}}\rvert =
2^{n-1}(2+n)\big(\sigma^2\rho^{(4)}_{0}\big)^{{n(n+1)}/{2}}.
\end{equation}
Together with \reflem{lem:independence}, these results
demonstrate the following theorem:
\begin{thm} \label{thm:pdfEigenvalues} The probability
  distribution $p_{\rand{L}_{\mathbf x}^{n}}$ of the eigenvalues
  of $\partial^2\rand{F}_{\mathbf x}^{n}$ is
  \begin{align}
    p_{\rand{L}_{\mathbf x}^{n}}(\boldsymbol{\lambda}) & =
    \frac{2^{(2-7n-n^2)/4}}
    {\sqrt{2+n}\big(\sigma^2\rho^{(4)}_{0}\big)^{{n(n+1)}/{4}}{\prod_{i=1}^{n}\Gamma(1+{i}/{2})}}\times\notag\\
    &\makebox[\widthof{\ensuremath{{} = {}}}]{}
    \prod_{i=1}^{n-1}
     \prod_{j=i+1}^{n}\lvert\lambda_{j}-\lambda_{i}\rvert
      \exp\bigg({-\frac{{\boldsymbol{\lambda}}^{\rm
        T}\tilde{\boldsymbol{\Sigma}}_{n}^{-1}{\boldsymbol{\lambda}}}{2}}\bigg).
    \label{eq:pdfEig}
  \end{align}
\end{thm}

Since $\rand{N}_{\mathbf x}$ in
\eqref{eq:simplifiedSpivakEquation} is a function of
$\partial\rand{F}_{\mathbf x}^{n}$,
\refthm{thm:covarianceMatrices}\eqref{eq:independence}
implies that $\partial^2\rand{F}_{\mathbf x}^{n}$ is independent
of $\rand{N}_{\mathbf x}$ for all $\mathbf x$. Therefore
$\partial\rand{F}_{\mathbf x}^{n-1} = \rand{N}_{\mathbf
  x}\partial\rand{F}_{\mathbf x}^{n}\rand{N}_{x}$ according to
\reflem{lem:invariance}.  \refthm[T]{thm:pdfEigenvalues} can
then be applied to obtain an expression for the probability
density of the eigenvalues of the numerator of
\eqref{eq:simplifiedSpivakEquation}, $p_{\rand{L}_{\mathbf
    x}^{n-1}}$. Using
\refthm{thm:covarianceMatrices}\eqref{eq:covDF}, we can
show that the denominator of
\eqref{eq:simplifiedSpivakEquation},
$\lVert\partial\rand{F}^{n}_{\mathbf x}\rVert$, is distributed
according to $\sigma(-\rho_{0}^{(0)})^{1/2}\rand{X}(n)$, where
$\rand{X}(n)$ follows a \emph{$\chi$-distribution} with $n$
degrees of freedom, and therefore its probability density
$p_{\lVert\partial\rand{F}^{n}_{\mathbf x}\rVert}$ is given by
\begin{equation}
  p_{\lVert\partial\rand{F}^{n}_{\mathbf x}\rVert}(u) =
  \frac{2u^{n-1}\exp[{u^2/(2\sigma^2\rho^{(2)}_{0})}]}{\big({-2}\sigma^2\rho^{(2)}_{0}\big)^{n/2}\Gamma(n/2)},
  \label{eq:chi}
\end{equation}
We can now prove our main result:
\begin{thm} \label{thm:pdfOfCurvatures} Let
  $\rand{K}$ be as in
  \eqref{eq:simplifiedSpivakEquation}. Then
  \begin{align}
    p_{\rand{K}_{\mathbf x}}(\boldsymbol{\kappa}) & = 
    \frac{2^{({n^2-7n+8})/{4}}\Gamma[n(n+1)/{4}]}
    {\sqrt{1+n}\,\Gamma({n}/{2})\prod_{i=1}^{n-1}\Gamma(1+{i}/{2})}\times\notag\\
    &\makebox[\widthof{\ensuremath{{} = {}}}]{}
    \frac{\alpha^{n(n-1)/{4}}\prod_{i=1}^{n-2}\prod_{j=i+1}^{n-1}\lvert\kappa_{j}-\kappa_{i}\rvert}
    {\big\{\alpha[\sum_{i=1}^{n-1}\kappa_{i}^2-\frac{1}{n+1}(\sum_{i=1}^{n-1}\kappa_{i})^2]+1\big\}^\frac{n^2+n}{4}},
    \label{eq:pdfOfCurvatures}
  \end{align}
  where $\alpha={-\rho^{(2)}_{0}}/({2\rho^{(4)}_{0}})$.
\end{thm}
\begin{proof}
  Since ${\partial\rand{F}_{\mathbf x}^{n}}$ and
  ${\partial^2\rand{F}_{\mathbf x}^{n-1}}$ are independent, so
  will be ${\lVert\partial\rand{F}_{\mathbf x}^{n}\rVert}$ and
  ${\rand{L}_{\mathbf x}^{n-1}}$. Using
  \eqref{eq:simplifiedSpivakEquation}, we have.
  \begin{equation}
    p_{\rand{K}_{\mathbf x}}(\boldsymbol{\kappa}) = 
    \int_{0}^{\infty}u^{n-1}p_{\rand{L}_{\mathbf x}^{n-1}}(\boldsymbol{\kappa}u)
    p_{\lVert\partial\rand{F}_{\mathbf x}^{n}\rVert}(u)\,du.\label{eq:pdfOfRatio}
  \end{equation}
  Substituting \eqref{eq:pdfEig} and \eqref{eq:chi} in
  \eqref{eq:pdfOfRatio}, we obtain \eqref{eq:pdfOfCurvatures}.
\end{proof}

\begin{acknowledgments}
  The authors would like to thank Robert Adler for his
  insightful comments.  This work was supported by the DOD
  and the Medical University of South Carolina under DOD
  Grant No.\ W81XWH-05-1-0378. Its contents are solely the
  responsibility of the authors and do not necessarily
  represent the official views of the Department of Defense
  or the Medical University of South Carolina.
\end{acknowledgments}

\appendix*
\section{Differentiability of the Autocorrelation Function\label{apen:differentiability}}
Correlation functions are characterized by the
Wiener-Khintchine theorem, a simplified version off which,
shown below, is quoted verbatim from \cite{Abrahamsen1997a}:
\begin{thm} \label{theo:Wiener-Khintchine} A real function
  $\rho(\lVert{\mathbf s}\rVert)$ on $\Real[n]$ is a correlation function if
  and only if it can be represented in the form 
\begin{equation}
\rho(\lVert{\mathbf s}\rVert) =
2^{(d-2)/2}\Gamma(d/2)\int_{0}^{\infty}\frac{J_{(d-2)/2}(k{s})}{(k{s})^{(d-2)/2}}\,d\Phi(k),
\end{equation}
where the function $\Phi(k)$ on $\Real$ has the properties
of a distribution function and $J_{\nu}$ is a Bessel
functions of the first kind and order $\nu$.
\end{thm}

\begin{lem} \label{lem:differentiability}
  Let the $i$-th moment of the distribution $\Phi(k)$ in
  \refthm{theo:Wiener-Khintchine} be defined. Then, the $i$-th derivative of
  $r(\lVert{\mathbf s}\rVert)$, $r^{(i)}(\lVert{\mathbf s}\rVert)$, exists and is given by
  \begin{equation}
    r^{(i)}(\lVert{\mathbf s}\rVert) =
    2^{(d-2)/2}\Gamma(d/2)\int_{0}^{\infty}k^{i}\frac{J_{(d-2)/2}(k{s})}{(k{s})^{(d-2)/2}}\,d\Phi(k).
  \end{equation}
\end{lem}

\begin{proof}
Define the operator $D_{i}$ acting on a function $f(u)$ as
\begin{equation}
D_{i}[f(u)] = \left(\frac{1}{u}\frac{d}{du}\right)_{i}[f(u)]
\end{equation}
where the term in the right-hand side is recursively defined as
\begin{align}
  \left(\frac{1}{u}\frac{d}{du}\right)_{1}[f(u)] & = \frac{1}{u}\frac{df(u)}{du}\\
  \left(\frac{1}{u}\frac{d}{du}\right)_{i}[f(u)] & =
  \left(\frac{1}{u}\frac{d}{du}\right)\left[\left(\frac{1}{u}\frac{d}{du}\right)_{i-1}[f(u)]\right].
\end{align}
It can be shown by induction that 
\begin{equation}
  r^{(i)}(\lVert{\mathbf s}\rVert) = 
  \left.\left(\frac{1}{u}\frac{d}{du}\right)_{i}[\rho(u)]\right|_{u=\sqrt{\lVert{\mathbf s}\rVert}}.
\end{equation}
The operator $D_{i}$ and the integral in \refthm{theo:Wiener-Khintchine} can be
interchanged, since the functions and the measure $d\Phi(k)$ involved satisfy
the conditions of Lebesgue's dominated convergence theorem. The identity
\begin{equation}
D_{i}\left[\frac{J_{\nu}(u)}{u^\nu}\right] = (-1)^{i}\frac{J_{\nu+i}(u)}{u^{\nu+i}},
\end{equation}
found in \cite{Abramowitz1972}, completes the proof.
\end{proof}

\bibliography{./paulo}
\end{document}